\documentclass[12pt]{article}
\usepackage{amssymb,amsfonts}
\usepackage{latexsym}
\usepackage{graphics}

\newfont{\DamirFont}{arxi}
\renewcommand{\footnote}[1]{%
\def\thefootnote{\arabic{footnote})}
\footnotemark%
\footnotetext{#1}}

\newfont{\blackboard}{msbm10 scaled\magstep2}
\newfont{\AMSSymbols}{msam10 }
\newcommand{\Z}{\mbox{\blackboard\symbol{"5A}}}

\def\Res{\mathop{\rm Res}}
\newcommand{\io}{[\hspace{-1pt}[}
\newcommand{\ic}{]\hspace{-1pt}]}
\newcommand{\fo}{\{\!\mid\!}
\newcommand{\fc}{\!\mid\!\}}

\newcommand{\bfalpha}{\mbox{\boldmath $\alpha$}}
\newcommand{\bsigma}{\mbox{\boldmath $\sigma$}}
\newcommand{\sgn}{{\rm sgn}}
\newcommand{\mod}{{\rm mod}}
\newcommand{\Sum}[1]{\displaystyle{\sum_{#1} \kern-1.40em \int}}

\begin{document}

\begin{center}
{\Large\bf
Self-Adjointness of the Dirac Hamiltonian and Vacuum
Quantum Numbers Induced by a Singular External Field}\\
\vspace{2mm}
{\bf
(Physics of Atomic Nuclei Vol.60: pp.2102-2113, 1997; \\
(Errata) Vol.62: p.1084, 1999.)}\\
\vspace{2mm}
{\bf Yu. A. Sitenko}\\
\vspace{1mm}
{Bogolyubov Institute for Theoretical Physics,\\
National Academy of Sciences of Ukraine, \\
Metrologicheskaya ul. 14b, Kiev, 252143 Ukraine}

\end{center}

Effects of fermion-vacuum polarization by a singular configuration of an 
external static vector field are considered in (2 + 1)-dimensional spacetime. 
Expressions for the induced vacuum charge and magnetic flux are obtained.

\section{Introduction}

\ \ \ Effects of singular external fields (zero-range potentials)
are comprehensively studied in quantum mechanics
(see \cite{Alb} and references therein). In this study, we
consider the effect of singular external fields on the fermion
vacuum in quantum field theory. In contrast to the
Schrodinger operator, the Dirac operator may be free of
an explicit delta-function singularity; nonetheless, the
problem of self-adjoint extension arises in both cases,
albeit for different reasons (see, for example \cite{Jac}). It is
well known that a singular magnetic-monopole external field leads to a $\Theta$
vacuum violating CP symmetry \cite{Gol,Cal,Wit,Gro,Yam}. It will be shown
below (see also \cite{Sit96}) that a singular magnetic-string external
field leads to a $\Theta$ vacuum violating C symmetry.

Our analysis will be based on representing the second-quantized
fermion-field operator in the form
\begin{equation}
\Psi({\bf x},t)=\Sum{E>0}
\quad e^{-iEt}\psi_{E}({\bf x})a_{E}+\Sum{E<0}
\quad e^{-iEt} \psi_{E}({\bf x})b_{E}^{\dag}.
\end{equation}
Here, the symbol $\Sum{E}$ denotes summation over the discrete
spectrum of the energy $E$ and integration (with
some measure) over its continuum spectrum; $a^{\dag}_{E}$ and $a_{E}$
($b^{\dag}_{E}$ and $b_{E}$) are the fermion (antifermion) creation and
annihilation operators satisfying anticommutation relations;
and $\psi_{E}({\bf x})$ is a solution to the equation
\begin{equation}
H\psi_{E}({\bf x}) =E\psi_{E}({\bf x}),
\end{equation}
where
\begin{equation}
H =-i {\bfalpha} \left[ \frac{\partial}{\partial{\bf x}}-i {\bf V}
({\bf x}) \right] + \beta m
\end{equation}
is the Dirac Hamiltonian in a static external vector field
${\bf V}({\bf x})$. If the condition
\begin{equation}
\int_{X}[{\tilde
{\psi}}^{\dag}(H\psi)-(H^{\dag}\tilde {\psi})^{\dag}\psi\rbrack \, d\Omega=0 ,
\end{equation}
where $d\Omega$ is the volume element of a spatial region $X$,
is met, the Hamiltonian $H$ is a Hermitian (symmetric)
operator acting in the space of functions defined on $X$.
If, in addition, the spaces of the functions $\psi$ and $\tilde{\psi}$
coincide, the Hamiltonian $H$ is a self-adjoint operator.

The integral on the left-hand side of (4) can be
reduced to an integral over the surface $\partial X$ bounding the
spatial region $X$. As a result, relation (4) takes the form
 \begin{equation}
-i\int_{\partial X}{\tilde{\psi}}^{\dag}{\bfalpha}\psi \cdot
d{\bsigma}=0,
\end{equation}
where $d{\bsigma}$ is an oriented element of the surface $\partial X$,
the normal to this element being directed outside the
region $X$.

A standard procedure leads to the expressions for
the vacuum-charge density,
\begin{equation}
\rho({\bf
x})=-\frac{1}{2}\Sum{E} \, \sgn(E)
\psi^{\dag}_{E}
({\bf x})\psi_{E}({\bf x}),
\end{equation}
and for the vacuum current,
\begin{equation}
{\bf j}({\bf
x})=-\frac{1}{2}\Sum{E} \, \sgn(E)\psi^{\dag}_{E}
({\bf x}) {\bfalpha} \psi_{E}({\bf x}),
\end{equation}
where
\begin {displaymath}
\sgn(u)=\biggl\lbrace{\quad 1,u>0\atop -1,u<0}.
\end{displaymath}
Performing integration in (6) over the spatial region $X$,
we obtain the total vacuum charge
\begin{equation}
Q^{(I)}=\int_{X}\rho\,d\Omega.
\end{equation}
As to a global quantity associated with vacuum current
(7), it follows from the Maxwell equation
\begin{equation}
\frac{1}{e^{2}}{\frac{\partial}{\partial {\bf x}}}\times{\bf B}_{(I)}({\bf
x})={\bf j}({\bf x})
\end{equation}
($e$ is the coupling constant) that a magnetic field is
induced in the vacuum. This magnetic field is characterized
by the field strength
\begin{equation}
{\bf B}_{(I)}({\bf x})=e^{2}\int_{{\bf x}_{(\infty)}}^{{\bf x}}{\bf j}({\bf
x})\times d{\bf x}\qquad
\left({\bf j}({\bf x}_{(\infty)})={\bf B}_{(I)}({\bf
x}_{(\infty)})=0\right)
\end{equation}
and by the total flux (in 2$\pi$ units)
\begin{equation}
\Phi^{(I)}=\frac{1}{2\pi}\int{\bf B}_{(I)}\cdot  d{\bsigma},
\end{equation}
where $d{\bsigma}$ is an oriented element of the surface orthogonal
to the lines of force of the magnetic field in the
spatial region $X$.

In this study, we consider a second-quantized fermion
field in an external field generated by a source in
the form of a singular magnetic string. If we direct the
coordinate $x^{3}$ axis along the string, the strength of the
string magnetic field is given by
\begin{equation}
B^{3}({\bf x}) = 2\pi\Phi^{(0)}\delta({\bf x}).
\end{equation}
By ${\bf x}$, we will henceforth mean a two-dimensional vector
in the plane orthogonal to the string axis [${\bf x}=(x^{1} , x^{2})$] ;
the parameter $\Phi^{(0)}$ is the total magnetic flux (in 2$\pi$
units) of a string. It is natural to choose the gauge
\begin{equation}
V_{3}=0 .
\end{equation}
The two-dimensional vector potential ${\bf V} = (V_{1} , V_{2})$ can
then be defined as
\begin{equation}
{\bf x}\cdot{\bf V}({\bf x}) = 0 ,
\quad{\bf x}\times{\bf V}({\bf x}) = \Phi^{(0)} .
\end{equation}
That the potential ${\bf V}({\bf x})$ is indeterminate at the point ${\bf x} = 0$
of the plane is associated with the delta-function singularity in the
strength $B^{3}$ (12) at this point. On the plane orthogonal to the string
axis, the Dirac Hamiltonian has the form
\begin{equation}
H =
-i\alpha^{r}\partial_{r} -
ir^{-1}\alpha^{\varphi}(\partial_{\varphi}-i\Phi^{(0)}) + \beta m ,
\end{equation}
where
\begin{equation}
\alpha^{r} = \alpha^{1}\cos\varphi + \alpha^{2}\sin\varphi,
\quad\alpha^{\varphi} = -\alpha^{1}\sin\varphi + \alpha^{2}\cos\varphi ,
\end{equation}
and where we introduced the polar coordinates
\begin{displaymath}
r = [(x^{1})^{2}+(x^{2})^{2}]^{\frac{1}{2}},\quad
\varphi = \arctan \left(\frac{x^{2}}{x^{1}}\right).
\end{displaymath}

This article is organized as follows. In Section 2, we
determine the complete system of solutions to the Dirac
equation in the field of a singular magnetic string. In
Section 3, we consider the vacuum charge induced on
the plane orthogonal to the string axis. In Section 4, we
analyze the vacuum magnetic flux through this plane.
The results obtained in this study are discussed in Section 5.
Some technical details concerning the derivation
of basic relations are described in Appendices A and B.

\section{Solving the Dirac equation in the field of a singular 
magnetic string}

\ \ \ It is well known that, in (2 + 1)-dimensional space-time,
the Clifford algebra does not have a faithful irreducible
representation; instead, it has two nonequivalent
irreducible representations. Accordingly, the matrices
$\alpha^{1} , \alpha^{2} $ and $\beta$ admit the following realizations in
terms of square rank-two matrices:
\begin{equation}
\alpha^{1} = -\sigma^2,\qquad \beta = \sigma^3,
\qquad\alpha^{2} =s\sigma^1,\qquad s = \pm 1.
\end{equation}
Here, $\sigma^{1} , \sigma^{2}$  and $\sigma^{3}$
are the Pauli matrices, and the two
possible values of the parameter $s$ correspond to the two
nonequivalent representations.

A solution to the time-independent Dirac equation
(2) with the Hamiltonian $H$ in the form (15) is given by
\begin{equation}
\psi_{E}({\bf x}) = \sum_{n \in \Z}<{\bf x}|E , n> ,
\end{equation}
where
\begin{equation}
<{\bf x}|E , n> = \left( \begin{array}{ccc}
f_{n}(r , E)e^{i n \varphi}\\
g_{n}(r , E)e^{i(n+s)\varphi}\\
\end{array} \right) ,
\end{equation}
and $\Z$ is the set of integers. The radial wave functions $f_{n}$ and $g_{n}$
satisfy the system of equations
\begin{equation}
(- \partial_{r} + r^{-1} s \lambda ) f_{n} = (E+m)g_{n} ,\qquad
[ \partial_{r} + r^{-1} (s \lambda + 1 )]g_{n} = (E-m)f_{n} ,
\end{equation}
where $\lambda = n-\Phi^{(0)}$.
In the case of $\Phi^{(0)}\not= n'$ (where $ n' \in \Z $),
two linearly independent solutions to (20) that correspond 
to the continuous spectrum, $|E| > |m|$, \footnote{
It should be noted that, in (2 + 1)-dimensional spacetime, as well
as in any spacetime having an odd number of dimensions, the
parameter $m$ appearing in expression (3) tor the Hamiltonian can
take both positive and negative values.} can be
represented as
\begin{equation}
\left( \begin{array}{ccc}
{f_{n}}^{(\pm)}(r , E)\\
{g_{n}}^{(\pm)}(r , E)\\
\end{array} \right) =
\left( \begin{array}{ccc}
f^{(0)}(E) J_{\pm s \lambda}(kr)\\
\pm g^{(0)}(E) J_{\pm(s \lambda+1)}(kr)\\
\end{array} \right) ,
\end{equation}
where $k=\sqrt{E^{2} - m^{2}}$, and $J_{\mu}(z)$ is the Bessel function of
order $\mu$. It can be seen from (21) that a solution that is
regular at the point $r = 0$ can be chosen for all modes
with the exception of that which corresponds to $n = n_{0}$,
where $n_{0}$ is determined from the condition
\begin{equation}
-1 < s{\lambda}_{0} < 0 .
\end{equation}
For this mode, either the upper or the lower component
of the spinor in (19) -- depending on the choice of the
plus or minus sign in (21) -- proves to be irregular at $r=0$,
although it is square-integrable. It is also obvious
that, for $\Phi^{(0)} = n'$ (where $ n' \in \Z)$), a solution that is regular
at the point $r = 0$ can be chosen for all modes.
Let us introduce the quantity
\begin{equation}
F = \frac{1}{2} + s\Bigl( \fo\Phi^{(0)} \fc - \frac{1}{2}
\Bigr).
\end{equation}
Here, $\fo u\fc$ stands for the fractional part of the quantity
$u$ -- that is $\fo u \fc = u - \io u\ic$ \\
($0\leq \fo u \fc<1$),
where $\io u\ic$ is the integral part of $u$ (the closest integer to $u$ from
below or the integer equal to $u$ if it is integral itself).
Taking into account the relations
\begin{equation}
n_{0} = \io\Phi^{(0)}\ic + \frac{1}{2} -
\frac{1}{2}s,\qquad s{\lambda}_{0} = -F,
\end{equation}
and the condition of orthonormality for states of the
continuous spectrum for $\sgn(E)=\sgn(E')$,
\begin{equation}
\int {d}^{2}x< E , n | {\bf x}><
{\bf x} | E' , n' > = \frac{\delta(k - k')}{\sqrt{k k'}}{\delta}_{n n'} ,
\end{equation}
we obtain
\begin{equation} \left( \begin{array}{ccc}
f_{n}\\ g_{n}\\ \end{array} \right) = \frac{1}{2{\sqrt{\pi}}} \left(
\begin{array}{ccc} {\sqrt{1 + m{E}^{-1}}} J_{l-F}(kr)\\ \sgn(E){\sqrt{1 -
m{E}^{-1}}} J_{l+1-F}(kr)\\ \end{array} \right) , l = s\Bigl( n -
\io\Phi^{(0)}\ic - \frac{1}{2} \Bigr) + \frac{1}{2} , \end{equation}
for the regular modes with $s\lambda > s{\lambda}_{0} (l \geq 1)$;
\begin{equation}
\left( \begin{array}{ccc} f_{n}\\ g_{n}\\ \end{array} \right) =
\frac{1}{2{\sqrt{\pi}}} \left( \begin{array}{ccc} {\sqrt{1 + m{E}^{-1}}} J_{
l' + F}(kr)\\ -\sgn(E){\sqrt{1 - m{E}^{-1}}} J_{ l' - 1 + F}(kr)\\
\end{array} \right),
 l' = -s\Bigl( n - \io\Phi^{(0)}\ic - \frac{1}{2} \Bigr) - \frac{1}{2},
\end{equation}
for the regular modes with $s\lambda < s{\lambda}_{0} ( l' \geq 1)$; and
\begin{equation}
\left( \begin{array}{ccc}
f_{{n}_{0}}\\
g_{{n}_{0}}\\
\end{array} \right) = \frac{1}{2 \sqrt{\pi(1 + {\sin 2\nu}\,{\cos F\pi})}}
\left( \begin{array}{ccc}
{\sqrt{1 + m{E}^{-1}}}[{\sin \nu} J_{-F}(kr) + {\cos \nu} J_{F}(kr)]\\
\sgn(E){\sqrt{1 - m{E}^{-1}}}[{\sin \nu} J_{1-F}(kr) - {\cos \nu}
J_{-1+F}(kr)\\
\end{array} \right) ,
\end{equation}
for the irregular mode ($s\lambda = s{\lambda}_{0}$). The parameter $\nu$ is
determined by the requirement that Hamiltonian (15) be a self-adjoint operator.
We are now going to consider this issue in some detail.

In the case being considered, the condition requiring
that Hamiltonian be a Hermitian operator (5) takes the
form
\begin{equation}
\left.\Bigl[ 2\pi r \sum\limits_{n\in \Z}({\tilde {f}_{n}}{g}_{n} -
{\tilde{g}_{n}} {f}_{n}) \Bigr]\right|_{r=0}^{r=\infty} = 0 .
\end{equation}
If the modes corresponding to $n \not= n_{0}$ are subjected to
the regularity condition at $r = 0$, Hamiltonian (15) as
defined on the space of these modes becomes a self-adjoint operator.
The mode with $n = n_{0}$ cannot be subjected to the regularity condition,
because we would then be obliged to discard solution (28), thereby spoiling
the completeness of the set of solutions to the Dirac
equation. Thus, there arises the problem of determining
the boundary condition for the irregular mode at $r = 0$ -- in 
other words, the problem of the self-adjoint extension of a Hermitian
operator (precisely for this mode).
This problem is solved with the aid of the Weyl-von
Neumann theory of self-adjoint operators (see, for
example, \cite{Alb,Reed}). Since the defect index of the operator
defined on the space of regular functions is $(1, 1)$ in the
case being considered, the self-adjoint extension represents a set
of operators that is parametrized with the aid of one real 
continuous variable $(\Theta)$, and the required boundary condition
for the irregular mode has the form
\footnote{For
$s = 1$ and $m > 0$, this boundary condition was obtained in \cite{Ger}.}
\begin{equation}
\lim_{r\to 0}{(|m|r)}^{F}\cos \Biggl( s\frac{\Theta}{2}
+\frac{\pi}{4} \Biggr) {f}_{n_{0}} = - \sgn(m) \lim_{r\to 0}{(|m|r)}^{1 -
F}\sin \Biggl( s\frac{\Theta}{2} + \frac{\pi}{4} \Biggr) {g}_{n_{0}}.
\end{equation}
Substituting the asymptotic form of solution (28) for
$r\to 0$ into (30), we arrive at
\begin{equation}
\tan\, \nu = \sgn(m\,E) {\sqrt{\frac{E - m}{E + m}}}{\Biggl( \frac{k}{|m|}
\Biggr)}^{2F - 1} A(F , \Theta) ,
\end{equation}
where
\begin{equation}
A ( F ,\Theta) = 2^{1-2F} \frac{\Gamma (1-F)}{\Gamma( F)} \tan \Biggl(
s\frac{\Theta}{2} + \frac{\pi}{4} \Biggr) ,
\end{equation}
and $\Gamma(z)$ is the Euler gamma function. Formulas (31)
and (32) establish the relation between the parameters
$\nu$ and $\Theta$. The boundary condition (30) results in that the
spectrum involves not only a continuum but also the bound state
\begin{equation}
\psi_{{BS}}({\bf{x}}) = \frac{\kappa}{\pi}{\sqrt{\frac{\sin F \pi}{1
+ ( 2F - 1) {m}^{-1}{E}_{{BS}}}}} \left( \begin{array}{ccc}
{\sqrt{1 + {m}^{-1}{E}_{{BS}}}} {K}_{F}(\kappa r) {e}^{i {n}_{0}
\varphi}\\
\sgn(m){\sqrt{1 - {m}^{-1}{E}_{{BS}}}} {K}_{1 - F}(\kappa r)
{e}^{i ({n}_{0} + s) \varphi}\\
\end{array} \right),
\end{equation}
where $\kappa = \sqrt{{m}^{2} - {E}^{2}_{{BS}}}$, $K_{\mu}(z)$
is the Macdonald function
of order $\mu$, and the bound-state energy $E_{{BS}}$ $(|E_{{BS}}| < |m| )$
is determined as a real-valued root to the algebraic
equation
\begin{equation}
{\sqrt{\frac{m + {E}_{BS}}{m - {E}_{BS}}}}{\Biggl( {\frac{\kappa}{|m|}}
\Biggr)}^{1 - 2F} = - A( F , \Theta).
\end{equation}
It is obvious that there is no bound state for
\begin{equation}
0 < A( F , \Theta ) < \infty
\end{equation}
and that there arises a bound state for
\begin{equation} -\infty <  A( F , \Theta ) < 0.
\end{equation}
The bound-state energy is zero, $E_{BS} = 0$,  at
\begin{equation} A( F , \Theta ) = -1.
\end{equation}
We also have
\begin{equation}
\sgn(E_{BS}) = \sgn(m)\quad ,  \quad-\infty < A( F , \Theta ) < -1 ,
\end{equation}
\begin{equation}
\sgn(E_{{BS}}) = - \sgn(m)\quad  ,  \quad -1 < A( F , \Theta ) < 0.
\end{equation}

Thus, we have constructed the complete system of
solutions to the Dirac equation in the field of a singular
magnetic string. It should be noted that for the case in
which $ s = 1$, $m > 0$ and $E > 0$, the results presented in
this section were first obtained in \cite{Ger}.

\section{Induced vacuum charge}

\ \ \ By using the explicit form of solutions to the Dirac
equation, we can find the vacuum-charge density averaged over all directions.
We have
\begin{equation}
\bar {\rho}(r) = \frac{1}{2 \pi}\int_{0}^{2 \pi}d \varphi \rho (\bf x) ,
\end{equation}
where $\rho(\bf x)$ is determined by relation (6).

For the contribution of the regular modes (26) and
(27) to the averaged vacuum charge $\bar{\rho}(r)$ (40), we
obtain
\begin{eqnarray}&&
\bar{\rho}_{{}_{\rm REG}}(r) = - \frac{1}{8 \pi} \int_{0}^{\infty}dk k \sum_{\sgn(E)}
|E|^{-1} \sum_{l = 1}^{\infty} \Bigl\{ (E + m) [ J^{2}_{l-F}(kr) +
J^{2}_{l+F}(kr) ] + \nonumber \\&&
+(E - m) [ J^{2}_{l+1-F}(kr) + J^{2}_{l-1+F}(kr)
] \Bigr\},
\end{eqnarray}
where we combined summation over $l$ and $l'$. In general,
the correct procedure should involve introducing
the regularizing factor $|E|^{-t} (t > 0)$ in (41) and going
over to the limit $t \rightarrow 0_{+}$ upon performing summation
and integration. However, the final result remains
unchanged if, instead of introducing a regularizing factor,
we perform first summation over the sign of the
energy and then integration with respect to $k$, the integral
of the contribution of each modes in (26) and (27)
with respect to $k$ being convergent and integration with
respect to $k$ being commutative with summation over $l$.
In our subsequent calculations, summation performed
first over the sign of the energy and then over $l$ is followed by
integration with respect to $k$.

Summation over the sign of energy and over $l$ in (41)
yields
\begin{equation}
\bar{\rho}_{{}_{\rm REG}}(r) = -\frac{m}{4\pi} \int_{0}^{\infty}dk
\frac{k}{\sqrt{k^{2} + m^{2}}} [J^{2}_{1-F}(kr) - J^{2}_{F}(kr)] .
\end{equation}
With the aid of the relation
\begin{displaymath}
\frac{1}{\sqrt{k^{2} + m^{2}}} = \frac{2}{\pi} \int_{0}^{\infty} du
\frac{1}{k^{2} + m^{2} + u^{2}}\, ,
\end{displaymath}
we can perform integration with respect to $k$. Following
the substitution $u = \sqrt{q^{2} - m^{2}}$, we eventually obtain
\begin{equation}
\bar {\rho}_{{}_{\rm REG}}(r) = - \frac{m}{2{\pi}^{2}} \int_{|m|}^{\infty} dq
\frac{q}{\sqrt{q^{2} - m^{2}}} [I_{1-F}(qr)K_{1-F}(qr) - I_{F}(qr)K_{F}(qr)],
\end{equation}
where $I_{\mu}(z)$ is the modified Bessel function of order $\mu$.
The expression coincident with (43) is obtained by
deforming the contour of integration in the complex
plane (see Appendix A).

An alternative representation of the contribution of
the regular modes to the averaged vacuum-charge density has the form
\begin{equation}
\bar{\rho}_{{}_{\rm REG}}(r) = - \frac{m\,r^{-1}}{(2\pi)^{\frac{3}{2}}} 
\int_{0}^{\infty} ds \,\exp (-s^{2} - \frac{m^{2}r^{2}}{2{s}^{2}})
[I_{1 - F}(s^{2}) - I_{F}(s^{2})],
\end{equation}
which can be obtained by using the relation
\begin{displaymath}
\frac{1}{\sqrt{k^{2} + m^{2}}}=\frac{2}{\sqrt{\pi}} \int_{0}^{\infty} du \, 
\exp [- u^{2}( k^{2} + m^{2} ) ].
\end{displaymath}

Taking into account (33)-(39), we find that the contribution
of the bound state is
\begin{eqnarray}&&
 \bar{\rho}_{{}_{\rm BS}}(r) = \sgn(m) \sgn(A + 1)[1 - \sgn(A)]
\frac{\sin F\pi}{{(2\pi)}^{2}} \frac{{\kappa}^{2}}{m + E_{{BS}}(2F- 1)}
 \times \nonumber
 \\&&
\times[(m + E_{{BS}}) K^{2}_{F}(\kappa r) + (m - E_{{BS}})
 K^{2}_{1 - F}(\kappa r)],
\end{eqnarray}
where the quantity $A$ is given by (32).

Taking into account (28) and (31), we obtain the
contribution of the irregular mode in the form
\begin{eqnarray}&&
\bar {\rho}_{{}_{\rm IRREG}}(r) = - \frac{1}{8\pi} \int_{0}^{\infty}
dk \frac{k}{\sqrt{k^{2} + m^{2}}} \{ A{k}^{2F}m{|m|}^{-2F}[ L_{(+)} +
 L_{(-)}]J^{2}_{-F}(kr) + \nonumber \\&&
+ A{k}^{-2(1-F)}m{|m|}^{-2F}[{(m - {\sqrt{k^{2} + m^{2}}})}^{2} L_{(+)}
+ {(m + {\sqrt{k^{2} + m^{2}}})}^{2} L_{(-)}]J^{2}_{1-F}(kr) + \nonumber \\&&
+ 2[(m + {\sqrt{k^{2} + m^{2}}}) L_{(+)} + (m - {\sqrt{k^{2} +
m^{2}}}) L_{(-)}]J_{-F}(kr)J_{F}(kr) + \nonumber      \\&&
+ 2[(m - {\sqrt{k^{2} +
m^{2}}}) L_{(+)} + (m + {\sqrt{k^{2} + m^{2}}})
L_{(-)}]J_{1-F}(kr)J_{-1+F}(kr) + \nonumber \\&&
+ {A}^{-1}k^{-2F}m^{-1}{|m|}^{2F}[{(m +
{\sqrt{k^{2} +m^{2}}})}^{2} L_{(+)} + {(m - {\sqrt{k^{2} +
m^{2}}})}^{2} L_{(-)}]J^{2}_{F}(kr) + \nonumber  \\&&
 + {A}^{-1}k^{2(1-F)}m^{-1}{|m|}^{2F}[ L_{(+)} +
L_{(-)}]J^{2}_{-1+F}(kr) \},
\end{eqnarray}
where summation over the sign of energy has been performed, and
\begin{equation}
L_{(\pm)} = {[A{k}^{-2(1-F)}m{|m|}^{-2F}(-m \pm {\sqrt{k^{2} +m^{2}}})
+ 2\cos F\pi +A^{-1}k^{-2F}m^{-1}{|m|}^{2F}(m \pm {\sqrt{k^{2} +
m^{2}}})]}^{-1}.
\end{equation}
In Appendix A, it is shown how expression (46) can be
transformed by deforming the contour in the complex
plane to arrive at the final form
\[
\bar{\rho}_{{}_{\rm IRREG}}(r) = \frac{m}{2{\pi}^{2}} \int_{|m|}^{\infty}
dq \frac{q}{\sqrt{q^{2} - m^{2}}}[I_{1-F}(qr)K_{1-F}(qr) - I_{F}(qr)
K_{F}(qr)] -
\frac{\sin F\pi}{{\pi}^{3}m} \int_{|m|}^{\infty}dq\times
\]
\[
\times
\frac{q^{3}}{\sqrt{q^{2}-m^{2}}} \frac{[1 + A \Bigl(
{\frac{q}{|m|}\Bigr)}^{-2(1-F)}] K^{2}_{F}(qr) - [1 + A^{-1}\Bigl(
{\frac{q}{|m|}\Bigr)}^{-2F}]K^{2}_{1-F}(qr)}{A {\Bigl(\frac{q}{|m|}}
{\Bigr)}^{2F} + 2 + A^{-1}{\Bigl( \frac{q}{|m|} \Bigr)}^{2(1-F)}}-
\]
\[
- \sgn(m)\sgn(A +1)[1 - \sgn(A)]\frac{\sin
F\pi}{(2\pi)^{2}}\frac{{\kappa}^{2}}{m + E_{{BS}}(2F-1)}\times
\]
\begin{equation}
\times[(m +
E_{{BS}})K^{2}_{F}(\kappa r) + (m - E_{{BS}})K^{2}_{1-F}(\kappa r)].
\end{equation}

Summing (43), (45), and (48), we find that the averaged
vacuum-charge density is given by
\[
\bar{\rho}(r) = -\frac{\sin F\pi}{{\pi}^{3}m} \int_{|m|}^{\infty}dq
\frac{q^{3}}{\sqrt{q^{2}-m^{2}}} \times
\]
\begin{equation}
\times\frac{[1 + A{\Bigl(\frac{q}{|m|}\Bigr)}^{-2(1-F)}]K^{2}_{F}(qr) - [1 +
A^{-1}{\Bigl(\frac{q}{|m|}\Bigr)}^{-2F}]K^{2}_{1-F}(qr)}
{A{ \Bigl(\frac{q}{|m|}\Bigr)}^{2F} + 2 + A^{-1}{ \Bigl(\frac{q}{|m|}
\Bigr)}^{2(1-F)}}.
\end{equation}
This expression tends to zero in proportion to $|m|^{\frac{1}{2}}
r^{-\frac{3}{2}}\exp (-2|m|r)$
for $r \to \infty$ and diverges in proportion to $|m|r^{-1}$ for $r \to 0$.
In the case of $\Theta \neq \frac{\pi}{2}(\mod\,\pi)$, integration of (49)
over the entire plane yields the expression for the vacuum charge induced by
a singular magnetic string. The resulting expression has the form
\begin{equation}
Q^{(I)} = -\frac{\sgn(m)}{2\pi} \int_{1}^{\infty}\frac{dv}{\sqrt{v-1}}
\frac{F[1 +Av^{-1+F}] - (1-F)[1 + A^{-1}v^{-F}]}
{Av^{F} + 2 + A^{-1}v^{1-F}}.
\end{equation}

It should be emphasized once again that relations
(49) and (50) hold only in the case of nonintegral values
of the string flux $(0 < F <1)$. For integral values of the
string flux $(F = 0)$, the density and the flux vanish
because all square-integrable modes are then regular
for $r \to 0$, and the latter case does not differ in the least
from the string-free case $({\Phi}^{(0)} = 0)$.

Expression (50) can be reduced to the form
\begin{equation}
\left.
Q^{(I)} = -\frac{1}{2} \sgn(m)\left[ F + \frac{2}{\pi}\arctan 
\left(\frac{1+Av^F}{\sqrt{v-1}}\right)\right|_{v=1}^{v=\infty}\right],
\end{equation}
whence we eventually obtain
\begin{equation}
Q^{(I)} =
\left\{\begin{array}{lr}
-\frac{1}{2}\sgn(m)[F-\sgn(A+1)],&0<F<\frac{1}{2}\\&\\
-\frac{1}{\pi}\,s\,\sgn(m)\arctan[\tan (\frac{\Theta}{2})], &
F=\frac{1}{2}\\&\\
\frac{1}{2}\sgn(m)[1-F-\sgn(A^{-1}+1)],&\frac{1}{2}<F<1
\end{array}\right..
\end{equation}
For the case of $F = \frac{1}{2}$, $s = 1$, and $m > 0$, the last relation
was obtained in \cite{Hou}. We also have
\begin{equation}
\lim_{F\to 0} Q^{(I)} = \frac{1}{2} \sgn(m),
\end{equation}
\begin{equation}
\lim_{F\to 1} Q^{(I)} = -\frac{1}{2}\sgn(m).
\end{equation}
In deriving the last two relations, we also considered
that
\begin{eqnarray}&&
\lim_{F\to 0} A = 0,\qquad \Theta \not=s\frac{\pi}{2}(\mod\,2\pi),\nonumber \\&&
\lim_{F\to 1}A^{-1}=0,\qquad\Theta \not=-s\frac{\pi}{2}(\mod\,2\pi).
\end{eqnarray}
From relations (53) and (54), it can be seen that the vacuum
charge as a function of the string flux undergoes
discontinuities at integral values of its argument.

This is confirmed in the case of $\Theta = \frac{\pi}{2}(\mod\,\pi)$ as
well. From (49), we can easily obtain
\begin{equation}
{\bar\rho}(r) = \left\{
\begin{array}{ll}
-\frac{\sin\,F\pi}{\pi^3}m\int_{|m|}^\infty
dq\,\frac{q}{\sqrt{q^2-m^2}}K^2_F(qr),&\Theta=s\frac{\pi}{2}
(\mod\,2\pi)\\ &\\
\frac{\sin\,F\pi}{\pi^3}m\int_{|m|}^\infty dq\,
\frac{q}{\sqrt{q^2-m^2}}K^2_{1-F}(qr),&\Theta=-s\frac{\pi}{2}(\mod\,2\pi)
\end{array}\right..
\end{equation}
Integrating this relation over the entire plane, we arrive at
\begin{equation}
Q^{(I)} = \biggl\lbrace{-\frac{1}{2}\sgn(m)F,\quad\Theta=s\frac{\pi}{2}(\mod\,
2\pi)\atop\frac{1}{2}\sgn(m)(1 - F),\quad\Theta=-s\frac{\pi}{2}(\mod\,2\pi)}.
\end{equation}

\section{Induced vacuum magnetic flux}

\ \ \ By using the explicit form of solutions to the Dirac
equation, we can find the vacuum current averaged over
all directions. We have
\begin{equation}
\bar{\bf{j}}(r) = \frac{1}{2\pi}\int_{0}^{2\pi}d\varphi\,\bf{j}(\bf{x}),
\end{equation}
where $\bf{j}(\bf{x})$ is given by (7). Having verified that the
radial component of the vacuum current vanishes,
$\bar{j}^{r}(r)=0$, we proceed to considering the angular component
$\bar{j}^{\varphi}(r)$.

For the contribution of the regular modes (26) and
(27) to $\bar{j}^{\varphi}(r)$, we obtain
\begin{equation}
\bar{j}^{\varphi}_{{}_{\rm REG}}(r) = -\frac{s}{2\pi}\int_{0}^{\infty}dk\,
\frac{k^{2}}{|E|}\sum_{l=1}^{\infty}[J_{l-F}(kr)J_{l+1-F}(kr) -
J_{l+F}(kr)J_{l-1+F}(kr)].
\end{equation}
Perfornung summation over $l$, we arrive at
\begin{equation}
\bar{j}^{\varphi}_{{}_{\rm REG}}(r) = -\frac{sr}{4\pi}\int_{0}^{\infty}dk\,
\frac{k^{3}}{\sqrt{k^{2}+m^{2}}}[J^{2}_{1-F}(kr) - J_{2-F}(kr)J_{-F}(kr) +
J_{1+F}(kr)J_{-1+F}(kr) - J^{2}_{F}(kr)].
\end{equation}
In the same way as that used to evaluate the contribution of the regular
modes to the averaged vacuum-charge density, the integral in (60) can be
reduced to an integral featuring modified Bessel functions in the corresponding
integral. The result has the form
\begin{eqnarray}&&
\bar{j}^{\varphi}_{{}_{\rm REG}}(r) = -\frac{s}{2{\pi}^{2}}\int_{|m|}^{\infty}dq\,
\frac{q^{2}}{\sqrt{q^{2}-m^{2}}}\{I_{1-F}(qr)K_{F}(qr) - I_{F}(qr)K_{1-F}(qr)
+\nonumber \\&&
+ \frac{2}{\pi}\sin F\pi[qr{K}^{2}_{1-F}(qr) - qr{K}^{2}_{F}(qr) +
2(F - \frac{1}{2})K_{F}(qr)K_{1-F}(qr)]\}.
\end{eqnarray}

With the aid of (33)-(39), we find that the contribution of the bound state
is given by
\begin{equation}
\bar{j}^{\varphi}_{{}_{\rm BS}}(r) = s\,\sgn(m)\sgn(A + 1)[1 - \sgn(A)]
\frac{\sin F\pi}{2{\pi}^{2}}\frac{{\kappa}^{3}}{m+E_{{BS}}(2F-1)}
K_{F}(\kappa r)K_{1-F}(\kappa r).
\end{equation}

Taking into account (28) and (31), we obtain the
contribution of the irregular mode in the form
\begin{eqnarray}&&
\bar{j}^{\varphi}_{{}_{\rm IRREG}}(r) =\frac{s}{4\pi}\int_{0}^{\infty}dk\,
\frac{k^{2}}{\sqrt{k^{2}+m^{2}}}\{A{k}^{-2(1-F)}m{|m|}^{-2F}[(m -
{\sqrt{k^{2}+m^{2}}})L_{(+)} +\nonumber\\&&+ (m +
{\sqrt{k^{2}+m^{2}}})L_{(-)}] J_{-F}(kr)J_{1-F}(kr)
+[L_{(+)} + L_{(-)}][J_{-F}(kr)J_{-1+F}(kr) - J_{F}(kr)J_{1-F}(kr)]+\nonumber
\\&& + A^{-1}k^{-2F}m^{-1}{|m|}^{2F}[(m + {\sqrt{k^{2}+m^{2}}})L_{(+)} +
(m - {\sqrt{k^{2}+m^{2}}})L_{(-)}]J_{F}(kr)J_{-1+F}(kr)\},
\end{eqnarray}
where the quantities $L_{(+)}$ and $L_{(-)}$ are given by (47). In
Appendix B, it is shown how expression (63) can be
transformed by deforming the contour of integration in
the complex plane to arrive at
\begin{eqnarray}&&
\bar{j}^{\varphi}_{{}_{\rm IRREG}}(r) =\frac{s}{2{\pi}^{2}}\int_{|m|}^{\infty}dq\,
\frac{q^{2}}{\sqrt{q^{2}-m^{2}}}[I_{1-F}(qr)K_{F}(qr) - I_{F}(qr)K_{1-F}(qr)]
- \nonumber \\&&
-\frac{s\,\sin\,F\pi}{{\pi}^{3}}\int_{|m|}^{\infty}dq\,\frac{q^{2}}{\sqrt{q^{2}
-m^{2}}}\frac{A{\bigl(\frac{q}{|m|}\bigr)}^{2F} - A^{-1}{\bigl(\frac{q}{|m|}
\bigr)}^{2(1-F)}}{A{\bigl(\frac{q}{|m|}\bigr)}^{2F} + 2 + A^{-1}{\bigl(
\frac{q}{|m|}\bigr)}^{2(1-F)}}
K_{F}(qr)K_{1-F}(qr) -\nonumber \\&&
- s\,\sgn(m)\sgn(A + 1)[1 - \sgn(A)]\frac{\sin\,F\pi}{2{\pi}^{2}}\,
\frac{{\kappa}^{3}}{m+E_{{BS}}(2F-1)}K_{F}(\kappa r)K_{1-F}(\kappa r).
\end{eqnarray}

Summing (61). (62). and (64). we find that the averaged vacuum current
has the form
\begin{eqnarray}&&
\bar{j}^{\varphi}(r) = -\frac{s\,\sin\,F\pi}{{\pi}^{3}}\int_{|m|}^{\infty}dq\,
\frac{q^{2}}{\sqrt{q^{2}-m^{2}}}\{qr[K^{2}_{1-F}(qr) - K^{2}_{F}(qr)]
+\nonumber \\&& +[2(F - \frac{1}{2})+
\frac{A{\bigl(\frac{q}{|m|}\bigr)}^{2F} -
A^{-1}{\bigl(\frac{q}{|m|}\bigr)}^{2(1-F)}}{A{\bigl(\frac{q}{|m|}\bigr)}^{2F}
+ 2 +
A^{-1}{\bigl(\frac{q}{|m|}\bigr)}^{2(1-F)}}]K_{F}(qr)K_{1-F}(qr)\}.
\end{eqnarray}
This expression tends to zero in proportion to
$|m|^{1/2}r^{-3/2}\exp(-2|m|r)$ for $r \to \infty$ and diverges in proportion
to $r^{-2}$ for $r\to 0$.

Averaging relation (10), where any point that is infinitely remote from
the string (that is, any point lying on the circle $r = \infty $) can be
taken for $\bf{x}_{(\infty)}$, we find that the averaged field strength
is given by
\begin{eqnarray}&&
\bar{B}_{(I)}^{3}(r) =
-\frac{s\,e^{2}\sin\,F\pi}{{\pi}^{3}}\int_{r}^{\infty}dr'\,\int_{|m|}^{\infty}dq
\,\frac{q^{2}}{\sqrt{q^{2}-m^{2}}}\{qr'[K^{2}_{1-F}(qr') -
K^{2}_{F}(qr')]+\nonumber \\&& +[2(F - \frac{1}{2}) +
\frac{A{\bigl(\frac{q}{|m|}\bigr)}^{2F} - A^{-1}{\bigl(\frac{q}{|m|}\bigr)}
^{2(1-F)}}{A{\bigl(\frac{q}{|m|}\bigr)}^{2F} + 2 + A^{-1}{\bigl(
\frac{q}{|m|}\bigr)}^{2(1-F)}}]K_{F}(qr')K_{1-F}(qr')\}.
\end{eqnarray}
For the total flux (11) of the vacuum magnetic field
induced by a singular magnetic string, we obtain
\begin{equation}
{\Phi}^{(I)} = -\frac{s\,e^{2}F(1-F)}{2\pi |m|}\left[\frac{1}{6}(F - \frac{1}{2}) +
\frac{1}{4\pi}\int_{1}^{\infty}\frac{dv}{v{\sqrt{v-1}}}\frac{A{v}^{F} -
A^{-1}v^{1-F}}{A{v}^{F} + 2 + A^{-1}v^{1-F}}\right].
\end{equation}
The coupling constant $e$ has dimensions of $\sqrt{|m|}$.
Expression (67) can also be recast into the form [compare with (50)]
\begin{equation}
{\Phi}^{(I)} = -\frac{s\,e^{2}F(1-F)}{12{\pi}^{2}|m|}\int_{1}^{\infty}
\frac{dv}{v{\sqrt{v-1}}}\frac{(1+F)(1 + A{v}^{F}) - (2-F)(1 +
A^{-1}v^{1-F})}{A{v}^{F} + 2 + A^{-1}v^{1-F}}.
\end{equation}
 In the case of half-integer values of the string flux, we
have
\begin{equation}
{\Phi}^{(I)} = -\frac{\,e^{2}}{8{\pi}^{2}|m|}\arctan[
\tan(\frac{\Theta}{2})],\qquad
F=\frac{1}{2}.  \end{equation}
It should be emphasized that, in contrast to the vacuum
charge, the vacuum magnetic flux is continuous at integral
values of the string flux.

\section{Discussion of results}

\ \ \ It has been shown that, on the plane orthogonal to
the singular magnetic string specified by (12), the fermion
vacuum is characterized by the quantum numbers
(52), (57), and (67), which depend on the parameter $\Theta$
of self-adjoint extension. In relation to the $\Theta$ vacuum
for the case of a monopole \cite{Gol, Cal,
Wit, Gro, Yam}, the $\Theta$ vacuum for the
case of a string possesses a richer structure [dependence on
${\Phi}^{(0)}$, $s$ and $\sgn(m)$)]. In all probability, this is
due to a nontrivial topology of the base space in the latter
case: ${\pi}_{1} = 0$ in the case of a space with a punctured
point, and ${\pi}_{1} = \Z$ in the case with a removed line (or in
the case of a plane with a punctured point), where ${\pi}_{1}$ is
the first homotopic group. It should be noted that the
vacuum charge changes sign and the vacuum magnetic flux 
remains unchanged under either of the substitutions
$s \to -s$ and $m \to -m$.

Let us compare expressions (52) and (57) obtained
here for the vacuum charge with the expression for the
vacuum charge induced by a regular configuration of an
external magnetic field. In the latter case, one has \cite{Nie}
\begin{equation}
Q^{(I)} = -\frac{1}{2}\,s\,\sgn(m)\,\Phi,\qquad \Phi = \frac{1}{2\pi}\int
d^{2}x\,{\tilde{B}}^{3}({\bf x}),
\end{equation}
where ${\tilde{B}}^{3}({\bf x})$ is a function that is continuous everywhere
with the exception of integrable singularities at
isolated points or on isolated lines. In contrast to (70),
expressions (52) and (57) are periodic in the flux of an
external magnetic field. This can be considered as a
manifestation of the Aharonov-Bohm effect \cite{Aha} in quantum 
field theory (see \cite{Sit90}). Since it is sometimes
stated that the vacuum charge is not periodic in ${\Phi}^{(0)}$
(see, for example, \cite{Fle, Par}), we will dwell on this point
at greater length.

Under charge conjugation,
\begin{equation}
C: \bf{V} \to -\bf{V}, \quad \psi \to {\sigma}_{1}{\psi}^{*},
\end{equation}
the charge operator and its vacuum expectation value,
as well as the vacuum magnetic flux, must change sign,
but this is not the case for expressions (52), (57), and
(67), because the boundary condition (30) violates
charge-conjugation symmetry. However, for a specific
choice of the parameter $\Theta$, this symmetry can be conserved.

In particular, the choice of
\begin{equation}
\left.\begin{array}{ll}
\Theta = s\frac{\pi}{2}(\mod\,2\pi),& s{\Phi}^{(0)} > 0\\
\Theta = -s\frac{\pi}{2}(\mod\,2\pi),& s{\Phi}^{(0)} < 0
\end{array}\right\}\quad
({\Phi}^{(0)}\not= n, \quad   n \in \Z),
\end{equation}
which corresponds to the boundary condition considered
in \cite{Alf, Gor}, leads to the expression (compare with
the results presented in \cite{Fle, Par})
\begin{equation}
Q^{(I)} =
\left\{\begin{array}{ll}
-\frac{1}{2}s\,\sgn(m)\fo{\Phi}^{(0)}\fc,&{\Phi}^
{(0)} > 0\\
\frac{1}{2}s\, \sgn(m)(1 - \fo{\Phi}^{(0)}\fc),&
{\Phi}^{(0)} < 0
\end{array}
\right\},\quad\quad
\fo\Phi^{(0)}\fc\not=0.
\end{equation}
This result changes sign under change conjugation, but
it is not periodic in ${\Phi}^{(0)}$.

The parameter $\Theta$ can be chosen in such a way as to
conserve both periodicity in ${\Phi}^{(0)}$ and the discrete symmetry (71).
This can be achieved, for example, by setting
\begin{equation}
\begin{array}{ll}
\Theta =s \frac{\pi}{2}(\mod\,2\pi),&-\frac{1}{2} < s
\bigl(\fo{\Phi}^{(0)}\fc - \frac{1}{2}\bigr) < 0\\
\Theta = 0\,(\mod\,2\pi),&\fo{\Phi}^{(0)}\fc =
\frac{1}{2}\\
\Theta = -s\frac{\pi}{2}(\mod\,2\pi),& 0 < s\bigl(\fo{\Phi}^{(0)}\fc -
\frac{1}{2}\bigr) < \frac{1}{2}
\end{array},
\end{equation}
which corresponds to the condition of minimal irregularity, i.e.,
to a radial wave function that diverges
for $r \to 0$ no faster than $r^{-p}$, where $p\leq\frac{1}{2}$. It is with
this boundary condition that the result reported in \cite{Sit90}
is recovered in the form
\begin{equation}
Q^{(I)} = \frac{1}{2}s\,\sgn(m)\left[\frac{1}{2}{\sgn}_{0}\bigl(\fo{\Phi}^{(0)}\fc -
\frac{1}{2}\bigr) - \fo{\Phi}^{(0)}\fc + \frac{1}{2}\right],
\end{equation}
where
\begin{displaymath}
{\sgn}_{0}(u) =
\left\{
\begin{array}{lc}
\sgn(u),&u\not= 0\\
0,&\,u = 0
\end{array}\right..
\end{displaymath}
It should be noted that expression (75) is continuous for
integral values of the string flux and displays discontinuities
at half-integer values.

Another choice of $\Theta$ that is also compatible both
with periodicity in ${\Phi}^{(0)}$ and with symmetry (71) is
\begin{equation}
\Theta = 0\,(\mod\,2\pi),\qquad 0 <\fo{\Phi}^{(0)}\fc < 1.
\end{equation}
We then arrive at the expression
\begin{equation}
Q^{(I)} = -\frac{1}{2}s\,\sgn(m)\left[\frac{1}{2}{\sgn}_{0}\bigl(\fo{\Phi}^{(0)}\fc -
\frac{1}{2}\bigr) + \fo{\Phi}^{(0)}\fc - \frac{1}{2}\right] ,
\end{equation}
which is discontinuous both at integral and half-integer
values of the string flux.

When the boundary condition (72) is used, expression (67)
for the vacuum magnetic flux takes the form (see also \cite{Fle})
\begin{equation}
{\Phi}^{(I)} =
\left\{
\begin{array}{ll}
-\frac{{e}^{2}}{12\pi|m|}\fo{\Phi}^{(0)}\fc
(1 - {\fo{\Phi}^{(0)}\fc}^{2}),& {\Phi}^{(0)} > 0\\
\frac{{e}^{2}}{12\pi|m|}(1 - \fo{\Phi}^{(0)}\fc)[1 - {(1 -
\fo{\Phi}^{(0)}\fc)}^{2}],& {\Phi}^{(0)} < 0
\end{array}\right..
\end{equation}
We also have
\begin{equation}
{\Phi}^{(I)} = \frac{\,e^{2}}{12\pi|m|}\fo{\Phi}^{(0)}\fc\Biggl(1 -
\fo{\Phi}^{(0)}\fc\Biggr) \left[\frac{3}{2}{\sgn}_{0}(\fo{\Phi}^{(0)}\fc - \frac{1}{2})
- \fo{\Phi}^{(0)}\fc + \frac{1}{2}\right] \end{equation}
for the boundary condition (74) and
\begin{equation} {\Phi}^{(I)} =
-\frac{s\,e^{2}F(1-F)}{2\pi |m|}\left[\frac{1}{6}(F - \frac{1}{2}) +
\frac{1}{4\pi}\int_{1}^{\infty}\frac{dv}{v{\sqrt{v-1}}}\frac{C_{F}v^{F} -
C_{F}^{-1}v^{1-F}}{C_{F}v^{F} + 2 + C_{F}^{-1}v^{1-F}}\right]
\end{equation}
for the boundary condition (76).

In the last expression, the quantity $F$ is given by
(23), and
\begin{equation}
C_{F} = 2^{1-2F}\frac{\Gamma(1-F)}{\Gamma(F)}.
\end{equation}

Expressions (79) and (80) are odd under charge conjugation and are
periodic in the string flux.

In conclusion, we note that the general form of the
boundary condition that conserves both $C$ symmetry
and periodicity in the string flux is given by
\begin{equation}
\begin{array}{lc}
\Theta =\Theta_C(\mod\,2\pi),&-\frac{1}{2}<s\bigl(\fo{\Phi}^{(0)}\fc
-\frac{1}{2}\bigr)<0\\&\\
\Theta =0\,(\mod\,2\pi),& \fo{\Phi}^{(0)}\fc=\frac{1}{2}\\&\\
\Theta=-\Theta_C(\mod\,2\pi),& 0<s\bigl(\fo{\Phi}^{(0)}\fc
-\frac{1}{2}\bigr)<\frac{1}{2}
\end{array}
,
\end{equation}
where $-\pi<\Theta_C\leq\pi$.

\section*{Acknowledgements}

I am grateful to S.A. Yushchenko for the contribution
that he made at initial stages of this study and to
L.D. Faddeev, A.V. Mishchenko, V.V. Skalozub, and
V.I. Skrypnik for discussion on results presented here.

This work was supported in part by the Ministry for
Science and Technologies of Ukraine and by the American Physical Society.

\section*{Appendix A}
\def\theequation{A.\arabic{equation}}
\setcounter{equation}{0}

On the basis of relations [see, for example, \cite{Abra})
\begin{eqnarray}&&\nonumber
J_{\mu}(iz) = \exp(\frac{i}{2}\mu\pi)I_{\mu}(z),\quad 
-\pi < \arg\,z\leq\frac{\pi}{2},\\&&  \nonumber
I_{\mu}(-z) = \exp(i\mu\pi)I_{\mu}(z),\,K_{\mu}(-z) = \exp(-i\mu\pi)K_{\mu}(z)-
i\pi\,I_{\mu}(z), \, -\pi < \arg\,z < 0,
\end{eqnarray}
we can easily obtain
\begin{eqnarray}&&\nonumber
J_{\mu}(kr)J_{\nu}(kr) = \frac{1}{2i\pi}\{\exp[\frac{i}{2}(\mu - \nu)\pi]
I_{\mu}(-ikr)K_{\nu}(-ikr) - \exp[\frac{i}{2}(\nu - \mu)\pi]I_{\mu}(ikr)
K_{\nu}(ikr) +\\&&
+ \exp[\frac{i}{2}(\nu - \mu)\pi]I_{\nu}(-ikr)K_{\mu}(-ikr) -
\exp[\frac{i}{2}(\mu - \nu)\pi]I_{\nu}(ikr)K_{\mu}(ikr)\}.
\end{eqnarray}
With the aid of (A.1), expression (46) can be recast into
the form
\begin{equation}
\bar{\rho}_{{}_{\rm IRREG}}(r) = \int_{C_{\mbox{\DamirFont\symbol{'001}}}}d\omega\,{\cal{F}}(\omega).
\end{equation}
Here, $\omega = k^{2}$ is the new variable of integration; the contour
$C_{\mbox{\DamirFont\symbol{'001}}}$ circumvents the real positive semiaxis
of the variable $\omega$, going along it at infinitely small distances
from below and above; and the integrand has the form
\begin{eqnarray}&&\nonumber
{\cal{F}}(\omega) =
\frac{i}{{(4\pi)}^{2}}\frac{1}{\varepsilon}\{A{\omega}^{F}m{|m|}^{-2F}[L_{(+)} +
L_{(-)}]I_{-F}(r{\sqrt{-\omega}})K_{F}(r{\sqrt{-\omega}}) +\\&& \nonumber
+ A{\omega}^{-1+F}m{|m|}^{-2F}[{(m - \varepsilon)}^{2}L_{(+)} + {(m +
\varepsilon)}^{2}L_{(-)}]I_{1-F}(r{\sqrt{-\omega}})K_{1-F}(r{\sqrt{-\omega}})
+\\&& \nonumber
+ {\omega}^{F}[(m + \varepsilon)L_{(+)} + (m -
\varepsilon)L_{(-)}]{(-\omega)}^{-F}I_{F}(r{\sqrt{
-\omega}})K_{F}(r{\sqrt{-\omega}}) +\\&& \nonumber
+ {\omega}^{-F}[(m + \varepsilon)L_{(+)} + (m - \varepsilon)L_{(-)}]{(-\omega)}^{F}
I_{-F}(r{\sqrt{-\omega}})K_{F}(r{\sqrt{-\omega}}) +\\&& \nonumber
+ {\omega}^{1-F}[(m - \varepsilon)L_{(+)} + (m + \varepsilon)L_{(-)}]{(-\omega)}^
{-1+F}I_{1-F}(r{\sqrt{-\omega}})K_{1-F}(r{\sqrt{-\omega}}) +\\&& \nonumber
+ {\omega}^{-1+F}[(m - \varepsilon)L_{(+)} + (m + \varepsilon)L_{(-)}]{(-\omega)}^
{1-F}I_{-1+F}(r{\sqrt{-\omega}})K_{1-F}(r{\sqrt{-\omega}}) + \\&& \nonumber
+ A^{-1}{\omega}^{-F}m^{-1}{|m|}^{2F}[{(m + \varepsilon)}^{2}L_{(+)} + {(m -
\varepsilon)}^{2}L_{(-)}]I_{F}(r{\sqrt{-\omega}})K_{F}(r{\sqrt{-\omega}}) +\\&&
+ A^{-1}{\omega}^{1-F}m^{-1}{|m|}^{2F}[L_{(+)} +
L_{(-)}]I_{-1+F}(r{\sqrt{-\omega}})K_{1-F}(r{\sqrt{-\omega}})\},\quad
\end{eqnarray}
where $\varepsilon = \sqrt{\omega + m^{2}}$. By continuously deforming the
contour of integration in the complex $\omega$ plane as is
shown in the Figure, we arrive at the relation
\begin{equation}
\int_{C_{\mbox{\DamirFont\symbol{'001}}}}d\omega\,{\cal F}(\omega) =
\int_{C_{\mbox{\DamirFont\symbol{'005}}}}d\omega\,{\cal F}(\omega) +
\int_{C_{\mbox{\DamirFont\symbol{'002}}}}d\omega\,{\cal F}(\omega) +
\int_{C_{\mbox{\DamirFont\symbol{'004}}}}d\omega\,{\cal F}(\omega) +
\int_{C_{\mbox{\DamirFont\symbol{'011}}}}d\omega\, {\cal F}(\omega).
\end{equation}
The integrals along the semicircles
$C_{\mbox{\DamirFont\symbol{'005}}}$ and $C_{\mbox{\DamirFont\symbol{'004}}}$
of infinite radii vanish, whereas the integral along the contour
circumventing the cut for $\omega < -{m}^{2}$ can be represented as
\begin{eqnarray}&&\nonumber
\int_{C_{\mbox{\DamirFont\symbol{'002}}}}d \omega {\cal F}(\omega) =
-\frac{1}{{(4\pi)}^{2}}\int_{m^{2}}^{\infty}\frac{du}{\sqrt{u - m^{2}}}\left(
A{u}^{F}m{|m|}^{-2F}\{e^{iF\pi}[R^{(+)}_{(+)} + R^{(+)}_{(-)}] + \right.
\\
&&\nonumber
+ e^{-iF\pi}[R^{(-)}_{(+)} + R^{(-)}_{(-)}]\}I_{-F}(r{\sqrt u})K_{F}(r{\sqrt
u}) - A{u}^{-1+F}m{|m|}^{-2F}\{e^{iF\pi}[{(m - i{\sqrt{u - m^{2}}})}^{2}
R^{(+)}_{(+)} +
\\
&&\nonumber
+ {(m + i{\sqrt{u - m^{2}}})}^{2}R^{(+)}_{(-)}] + e^{-iF\pi}[{(m - i{\sqrt
{u - m^{2}}})}^{2}R^{(-)}_{(+)}
+ {(m + i{\sqrt{u -m^{2}}})}^{2}R^{(-)}_{(-)}]\}\times
\\
&&\nonumber
\times I_{1-F}(r{\sqrt{u}})K_{1-F}(r{\sqrt{u}})
+\{e^{iF\pi}[(m + i{\sqrt{u - m^{2}}})R^{(+)}_{(+)}+
(m - i{\sqrt{u -
m^{2}}})R^{(+)}_{(-)}]+
\\
&&\nonumber
+ e^{-iF\pi}[(m + i{\sqrt{u - m^{2}}})R^{(-)}_{(+)} +
(m - i{\sqrt{u - m^{2}}})R^{(-)}_{(-)}]\}I_{F}(r{\sqrt{u}})K_{F}(r{\sqrt{u}})
 +
\\
&& \nonumber
+\{ e^{-iF\pi}[(m + i{\sqrt{u -m^{2}}})R^{(+)}_{(+)} + (m - i{\sqrt{u -
m^{2}}})R^{(+)}_{(-)}] + e^{iF\pi}[(m + i{\sqrt{u - m^{2}}})R^{(-)}_{(+)} +
\\
&&\nonumber
+ (m - i\sqrt{u -m^{2}})R^{(-)}_{(-)}]\}
I_{-F}(r{\sqrt{u}})K_{F}(r{\sqrt{u}})
- \{e^{-iF\pi}[(m
- i{\sqrt{u - m^{2}}})R^{(+)}_{(+)} +
\\
&&\nonumber
+ (m + i{\sqrt{u - m^{2}}})R^{(+)}_{(-)}]
+
e^{iF\pi}[(m - i{\sqrt{u - m^{2}}})R^{(-)}_{(+)}+
(m + i{\sqrt{u -m^{2}}})R^{(-)}_{(-)}]\}\times
\\
&&\nonumber
\times I_{1-F}(r{\sqrt{u}})K_{1-F}(r{\sqrt{u}}) -
\{e^{iF\pi}[(m - i{\sqrt{u - m^{2}}})R^{(+)}_{(+)}+
(m + i{\sqrt{u -
 m^{2}}}R^{(+)}_{(-)}]+
\\
&&\nonumber
+ e^{-iF\pi}[(m - i{\sqrt{u - m^{2}}})R^{(-)}_{(+)}
 + (m + i{\sqrt{u -
 m^{2}}})R^{(-)}_{(-)}]\}I_{-1+F}(r{\sqrt{u}})K_{1-F}(r{\sqrt{u}})+
\\
&&\nonumber
 + A^{-1}
u^{-F}m^{-1}{|m|}^{2F}\{e^{-iF\pi}[{(m + i{\sqrt{u - m^{2}}})}^{2}R^{(+)}_{(+)}
+ (m - i\sqrt{u - m^2})^2R^{(+)}_{(-)}] +
\\ &&\nonumber
+e^{iF\pi}[(m + i\sqrt{u
-m^2})^2R^{(-)}_{(+)} + (m - i\sqrt{u - m^2})^2R^{(-)}_{(-)}]\}
I_{F}(r\sqrt{u})K_{F}(r\sqrt{u}) -
\\ &&  
 - A^{-1}u^{1-F}m^{-1}{|m|}^{2F}\{e^{-iF\pi}[R^{(+)}_{(+)} + R^{(+)}_{(-)}]
 + e^{iF\pi}[R^{(-)}_{(+)} + R^{(-)}_{(-)}]\}I_{-1+F}(r{\sqrt{u}})K_{1-F}(r{
 \sqrt{u}})\biggr)
\end{eqnarray}
where
\begin{eqnarray}\nonumber
R^{(+)}_{(\pm)}&=&[Au^{-1+F}m{|m|}^{-2F}e^{iF\pi}(m\mp i\sqrt{u -m^{2}})
+2\cos F\pi+
\\
&&\nonumber
+ A^{-1}u^{-F}m^{-1}|m|^{2F}e^{-iF\pi}(m \pm i\sqrt{u -m^{2}})]^{-1},
\\
\nonumber
R^{(-)}_{(\pm)}& =& [Au^{-1+F}m|m|^{-2F}e^{-iF\pi}(m \mp i\sqrt{u - m^2})
+ 2\cos F\pi +
\\
&&
+ A^{-1}u^{-F}m^{-1}|m|^{2F}e^{iF\pi}(m\pm i\sqrt{u-m^2})]^{-1}.
\end{eqnarray}

Expression (A.5) can be reduced to the form:
\begin{eqnarray}&&\nonumber
\int_{C_{\mbox{\DamirFont\symbol{'002}}}}d\omega{\cal F}(\omega) = -\frac{1}{8{\pi}^{2}}\int_{m^{2}}^{\infty}
\frac{du}{\sqrt{u - m^{2}}}\biggl(\frac{1}{\pi}\sin F\pi\{Au^{F}m{|m|}^{-2F}[
e^{iF\pi}(R^{(+)}_{(+)} + R^{(+)}_{(-)}) +\\ \nonumber
&&+ e^{-iF\pi}(R^{(-)}_{(+)} +
R^{(-)}_{(-)})] +
(m + i{\sqrt{u -m^{2}}})(e^{-iF\pi}R^{(+)}_{(+)} + e^{iF\pi}R^{(-)}_{(+)})+
\\ \nonumber
&&
+ (m - i{\sqrt{u - m^{2}}})(e^{-iF\pi}R^{(+)}_{(-)} +
e^{iF\pi}R^{(-)}_{(-)})\}K^{2}_{F}(r{\sqrt{u}}) +
\\&&  \nonumber
 + 2m[I_{F}(r{\sqrt{u}})K_{F}(r{\sqrt{u}}) -
 I_{1-F}(r{\sqrt{u}})K_{1-F}(r{\sqrt{u}})] -\frac{1}{\pi}\sin F\pi\{(m
 -i{\sqrt{u -m^{2}}})(e^{iF\pi}R^{(+)}_{(+)} + \\&&   \nonumber
+ e^{-iF\pi}R^{(-)}_{(+)}) + (m + i{\sqrt{u -m^{2}}})(e^{iF\pi}R^{(+)}_{(-)}
+ e^{-iF\pi}R^{(-)}_{(-)}) +
A^{-1}u^{1-F}m^{-1}{|m|}^{2F}[e^{-iF\pi}(R^{(+)}_{(+)} +\\&&
+ R^{(+)}_{(-)}) + e^{iF\pi}(R^{(-)}_{(+)} +
R^{(-)}_{(-)})]\}K^{2}_{1-F}(r\sqrt{u})\biggr).
\end{eqnarray}
As the result of further simplifications, we will arrive at
the terms in relation (48) that are represented as integrals 
with respect to the variable $q = \sqrt{u}$.

It remained to consider the integral along the contour 
circumventing the pole of the function $\cal{F}(\omega)$. We
have
\begin{equation}
\int_{C_{\mbox{\DamirFont\symbol{'011}}}}d\omega{\cal F}(\omega) = 2\pi i
\Res_{\omega = -{\kappa}^{2}}{\cal F}(\omega),
\end{equation}
where ${\kappa}^{2} = m^{2} - {E_{{BS}}}^{2}$, and the quantity $E_{{BS}}$
is given by (34). Choosing the branch for fractional exponents
according to the prescription
\begin{equation}
(-\kappa^2)^{\mu} = {\kappa}^{2\mu}\exp(i\mu\pi),\quad 0 < \mu < 1,\qquad
\end{equation}
we arrive at
\begin{eqnarray}
& &\nonumber
\int_{C_{\mbox{\DamirFont\symbol{'011}}}} d\omega{\cal F}(\omega)=
-\frac{1}{8\pi|E_{{BS}}|}[A{\kappa}^{2F}m
{|m|}^{-2F}e^{iF\pi}I_{-F}(\kappa r)K_{F}(\kappa r) - \nonumber \\
&&  \nonumber
- A{\kappa}^{-2(1-F)}m{|m|}^{-2F}e^{iF\pi}{(m \mp |E_{{BS}}|)^{2}I_{1-F}
(\kappa r)K_{1-F}(\kappa r)} +\\&&  \nonumber
+ e^{iF\pi}(m \pm |E_{{BS}}|)I_{F}(\kappa r)K_{F}(\kappa r) + e^{-iF\pi}
(m \pm |E_{{BS}}|)I_{-F}(\kappa r)K_{F}(\kappa r) -\\&&  \nonumber
- e^{-iF\pi}(m \mp |E_{{BS}}|)I_{1-F}(\kappa r)K_{1-F}(\kappa r) -
e^{iF\pi}(m \mp |E_{{BS}}|)I_{-1+F}(\kappa r)K_{1-F}(\kappa r) +\\
&& \nonumber
+ A^{-1}{\kappa}^{-2F}m^{-1}{|m|}^{2F}e^{-iF\pi}{(m \pm |E_{{BS}}|)}^{2}I_{F}
(\kappa r)K_{F}(\kappa r) - \\&&  \nonumber
- A^{-1}{\kappa}^{2(1-F)}m^{-1}{|m|}^{2F}e^{-iF\pi}I_{-1+F}(\kappa r)K_{1-F}
(\kappa r)]\Res_{\omega = -{\kappa}^{2}} L_{(\pm)} = \\
&=& \frac{i{\sin}^{2}F\pi}{2{\pi}^{2}|E_{{BS}}|}[(m \pm |E_{{BS}}|)K^{2}_
{F}(\kappa r) + (m \mp |E_{{BS}}|)K^{2}_{1-F}(\kappa r)]\Res_{\omega =
-{\kappa}^{2}}L_{(\pm)},\quad   E_{{BS}}\gtrless 0.\,
\end{eqnarray}
Taking into account the relation
\begin{equation}
\Res_{\omega=-\kappa^2} L_{(\pm)}
=\frac{1}{i \sin F\pi}
\frac{|E_{BS}|\kappa^2}{|E_{BS}|(2F-1) \pm m}\;,\quad
E_{BS} \gtrless 0\,,
\label{A.11}
\end{equation}
we obtain
\begin{equation}
\int_{C_{\mbox{\DamirFont\symbol{'011}}}}\,d\omega\;{\cal F}(\omega)=
{\rm sgn}(E_{BS})\frac{\sin F\pi}{2\pi^2}
\frac{\kappa^2}{m+E_{BS}(2F-1)}\left[\left(m+E_{BS}\right)
K^2_{F}(\kappa r)+\left(m-E_{BS}\right)K^2_{1-F}(\kappa r)\right].
\label{A.12}
\end{equation}
Naturally, the same expression is obtained if the branch
for fractional exponents is chosen alternatively as
\begin{equation}
\left(-\kappa^2\right)^{\mu}=\kappa^{2\mu}\exp (-i\mu\pi)\,, \quad
0<\mu<1.
\label{A.13}
\end{equation}
With the aid of relations (35)-(39), expression (A.12) is
reduced to the form coincident with that of the last term
in (48).

It should be also noted that the above method can be
used to reduce expression (42) to the form (43), in
which case the integrand naturally does not have poles
on the segment $-m^2<k^2<0$.

\section*{Appendix B}

The contribution of the irregular mode to the averaged 
vacuum current can be found by a method that is
similar to that used to calculate the contribution of this
mode to the averaged density of the vacuum charge. In
the case being considered, the function ${\cal F}(\omega)$ has the
form
\begin{eqnarray}
& &{\cal F}(\omega)=\frac{s}{i(4\pi)^2}\frac{1}{\varepsilon}\left(A
\omega^{F}m|m|^{-2F}\left[(m-\varepsilon)L_{(+)}+
(m+\varepsilon)L_{(-)}\right](-\omega)^{-\frac{1}{2}}
\times \right. \nonumber \\
& & \times \left[I_{1-F}\left(r\sqrt{-\omega\,}\right)
K_{F}\left(r\sqrt{-\omega\,}\right)-
I_{-F}\left(r\sqrt{-\omega\,}\right)
K_{1-F}\left(r\sqrt{-\omega\,}\right)\right]-\left[L_{(+)}+
L_{(-)}\right]\times
\nonumber \\
& & \times \left\{\omega^{1-F}
\left(-\omega\right)^{-\frac{1}{2}+F}
\left[I_{1-F}\left(r\sqrt{-\omega\,}\right)
K_{F}\left(r\sqrt{-\omega\,}\right)-I_{-F}\left(r\sqrt{-\omega\,}\right)
K_{1-F}\left(r\sqrt{-\omega\,}\right)\right]+\right.
\nonumber \\
& & \left.+\omega^{F}\left(-\omega\right)^{\frac{1}{2}-F}
\left[I_{F}\left(r\sqrt{-\omega\,}\right)
K_{1-F}\left(r\sqrt{-\omega\,}\right)-
I_{-1+F}\left(r\sqrt{-\omega\,}\right)
K_{F}\left(r\sqrt{-\omega\,}\right)\right]\right\}+
\nonumber \\
& & +A^{-1}\omega^{1-F}m^{-1}|m|^{2F}
\left[(m+\varepsilon)L_{(+)}+
(m-\varepsilon)L_{(-)}\right](-\omega)^{-\frac{1}{2}}
\left[I_{F}\left(r\sqrt{-\omega\,}\right)
K_{1-F}\left(r\sqrt{-\omega\,}\right)-
\right. \nonumber \\
& & \left.
-I_{-1+F}\left(r\sqrt{-\omega\,}\right)
K_{F}\left(r\sqrt{-\omega\,}\right)\right]\Bigr)\;.
\label{B.1}
\end{eqnarray}
The integral along the contour circumventing the cut
for $\omega<-m^2$ (see Figure) can be reduced to the form
\begin{eqnarray}
& & \int_{C_{\mbox{\DamirFont\symbol{'002}}} }\,d\omega\;{\cal F}(\omega)=
\frac{s}{(4\pi)^2}\int\limits_{m^2}^{\infty}du\,\left(1-
\frac{m^2}{u}\right)^{-\frac{1}{2}}\left(A u^{-1+F}m|m|^{-2F}
\left\{ e^{iF\pi} \left[\left(m-i\sqrt{u-m^2\,}\right)R^{(+)}_{(+)}+
\right. \right. \right. \nonumber \\
& & \left. \left. +
\left(m+i\sqrt{u-m^2\,}\right)R^{(+)}_{(-)}\right]+
e^{-iF\pi} \left[\left(m-i\sqrt{u-m^2\,}\right)R^{(-)}_{(+)}+
\left(m+i\sqrt{u-m^2\,}\right)R^{(-)}_{(-)}\right]\right\}
\times \nonumber \\
& & \times
\left[I_{1-F}\left(r\sqrt{u}\right)K_{F}\left(r\sqrt{u}\right)-
I_{-F}\left(r\sqrt{u}\right)K_{1-F}\left(r\sqrt{u}\right)\right]+
\left\{ e^{-iF\pi} \left[R^{(+)}_{(+)}+
R^{(+)}_{(-)}\right]+
e^{iF\pi} \left[R^{(-)}_{(+)}+ \right. \right. \nonumber \\
& & \left. \left. +
R^{(-)}_{(-)}\right]\right\}
\left[I_{1-F}\left(r\sqrt{u}\right)K_{F}\left(r\sqrt{u}\right)-
I_{-F}\left(r\sqrt{u}\right)K_{1-F}\left(r\sqrt{u}\right)\right]-
\left\{ e^{iF\pi} \left[R^{(+)}_{(+)}+
R^{(+)}_{(-)}\right]+ \right. \nonumber \\
& & \left. + e^{-iF\pi}\left[R^{(-)}_{(+)}+
R^{(-)}_{(-)}\right]\right\}
\left[I_{F}\left(r\sqrt{u}\right)K_{1-F}\left(r\sqrt{u}\right)-
I_{-1+F}\left(r\sqrt{u}\right)K_{F}\left(r\sqrt{u}\right)\right]-
\nonumber \\
& & -A^{-1}u^{-F}m^{-1}|m|^{2F}\left\{ e^{-iF\pi}
\left[\left(m+i\sqrt{u-m^2\,}\right)R^{(+)}_{(+)}+
\left(m-i\sqrt{u-m^2\,}\right)R^{(+)}_{(-)}\right]+
\right. \nonumber \\
& &\left. +e^{iF\pi} \left[\left(m+
i\sqrt{u-m^2\,}\right)R^{(-)}_{(+)}+
\left(m-i\sqrt{u-m^2\,}\right)R^{(-)}_{(-)}\right]\right\}
\left[I_{F}\left(r\sqrt{u}\right)K_{1-F}\left(r\sqrt{u}\right)-
\right. \nonumber \\
& & \left. \left.
-I_{-1+F}\left(r\sqrt{u}\right)
K_{F}\left(r\sqrt{u}\right)\right]\right)\,,
\label{B.2}
\end{eqnarray}
where the quantity $R^{(\pm)}_{(\pm)}$ is given by (A.6). Simplifying
the last expression, we arrive at the terms in relation
(64) that are represented as integrals.

For the integral along the contour circumventing the
pole of the function ${\cal F}(\omega)$ (see Figure), we have
\begin{eqnarray}
& & \int\limits_{C_{\mbox{\DamirFont\symbol{'011}}} }\,d\omega\;
{\cal F}(\omega)=
\frac{s\,\kappa}{8\pi |E_{BS}|}\left\{ \left[A\kappa^{-2(1-F)}
m|m|^{-2F}e^{iF\pi}\left(m\mp |E_{BS}|\right)+e^{-iF\pi}\right]
\left[I_{1-F}\left(\kappa r\right)K_{F}\left(\kappa r\right)-
\right. \right. \nonumber \\
& & \left.
-I_{-F}\left(\kappa r\right)K_{1-F}\left(\kappa r\right)\right]-
\left[e^{iF\pi}+A^{-1}\kappa^{-2F}
m^{-1}|m|^{2F}e^{-iF\pi}\left(m\pm |E_{BS}|\right)\right]
\left[I_{F}\left(\kappa r\right)K_{1-F}\left(\kappa r\right)-
\right. \nonumber \\
& & \left. \left.
-I_{-1+F}\left(\kappa r\right)K_{F}\left(\kappa r\right)\right] \right\}
\Res_{\omega=-\kappa^2} L_{(\pm)}\,,
\quad E_{BS} \gtrless 0\,,
\label{B.3}
\end{eqnarray}
where we choose the branch $(-\kappa^2)^{\mu}$ according to
(A.9). Taking into account (A.11), we obtain
\begin{equation}
\int_{C_{\mbox{\DamirFont\symbol{'011}}}}\,d\omega{\cal F}\;(\omega)=
{\rm s\,sgn}(E_{BS})\frac{\sin F\pi}{\pi^2}
\frac{\kappa^3}{m+E_{BS}(2F-1)}K_{F}(\kappa r)K_{1-F}(\kappa r)\,.
\label{B.4}
\end{equation}
With the aid of (35)-(39), expression (B.4) is reduced
to the same form as that of the last term in (64).

\includegraphics{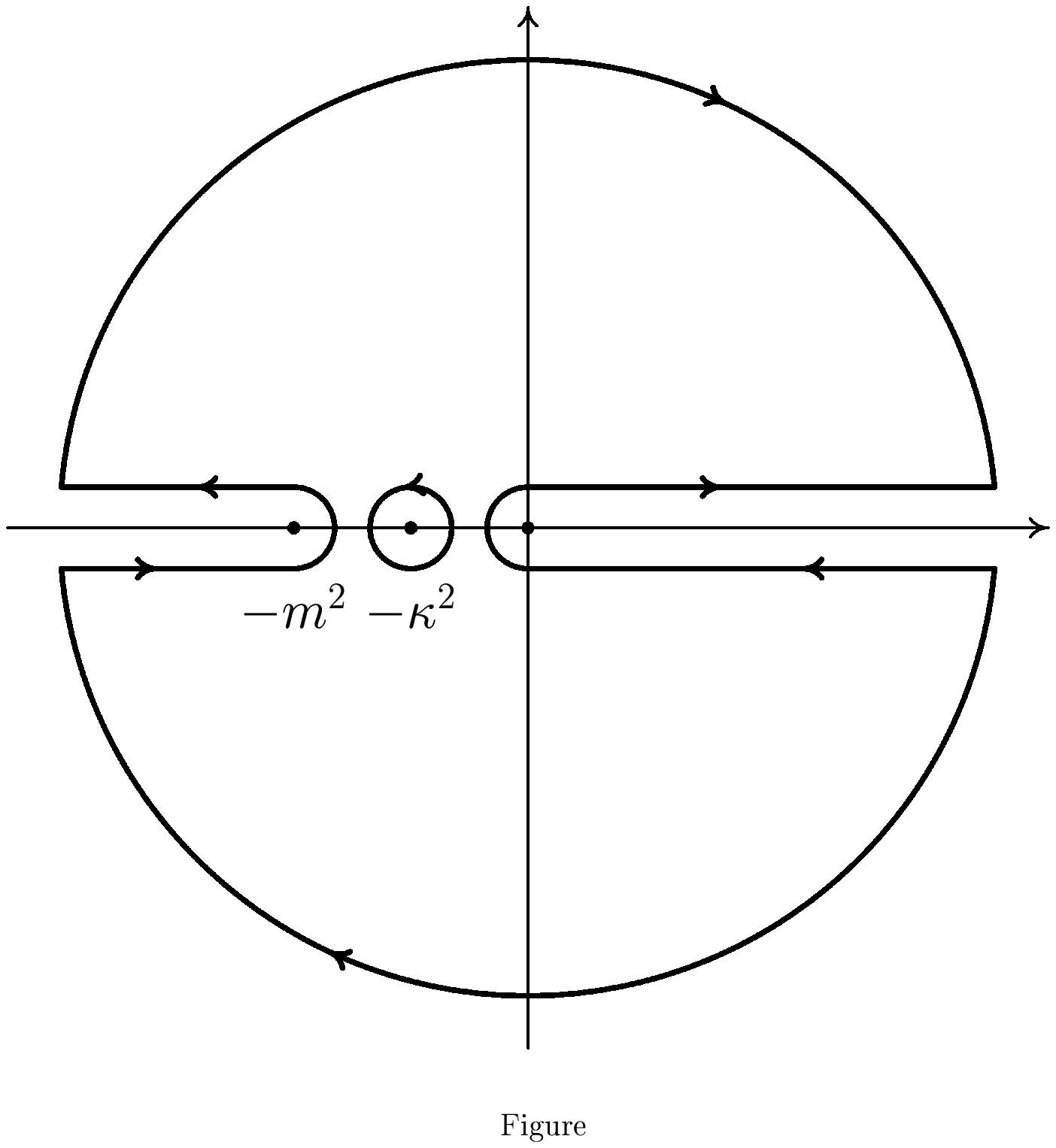}

\end{document}